\documentclass{jkas}

\def\year{2025} 
\def\volume{00} 
\def\issue{0} 
\def\beginpage{1} 
\def\received{2025 March 14} 
\def\accepted{2025 April 30} 
\date{Received \received; Accepted \accepted}

\newcommand{\micron}[1]{{#1} $\mu$m} 

\usepackage{textcomp}
\usepackage{chemformula}
\newcommand{\RomanNumeralCaps}[1]
{\MakeUppercase{\romannumeral #1}}
\title{%
Near- to mid-infrared spectroscopic study of ice analysis using the AKARI/IRC and Spitzer/IRS spectra
}

\author[1]{Jaeyeong Kim}{0000-0001-8604-2801}
\author[2,4]{Jeong-Eun Lee}{0000-0003-3119-2087}
\author[2]{Chul-Hwan Kim}{0000-0002-2523-3762}
\author[1]{Woong-Seob Jeong}{0000-0002-2770-808X}
\author[3]{Yao-Lun Yang}{0000-0001-8227-2816}

\affil[1]{Korea Astronomy and Space Science Institute, 776 Daedeok-daero, Yuseong-gu, Daejeon 34055, Republic of Korea}
\affil[2]{Department of Physics and Astronomy, SNU Astronomy Research Center, Seoul National University, 1 Gwanak-ro, Gwanak-gu, Seoul 08826, Republic of Korea}
\affil[3]{Star and Planet Formation Laboratory, RIKEN Cluster for Pioneering Research, Wako-shi, Saitama, 351-0106, Japan}
\affil[4]{\email{lee.jeongeun@snu.ac.kr}}

\def\corrauthor{%
J. Kim, \email{jaeyeong@kasi.re.kr}
}

\def\runningauthor{%
Kim et al.
}

\def\runningtitle{%
Infrared spectroscopic ice analysis
}

\def\keywords{%
Astrochemistry --- Spectroscopy: infrared: AKARI and Spitzer --- Protostars
}

\def\abstracttext{%
We present the combined 2.5$-$\micron{30} spectra of four protostars acquired with the infrared camera and the infrared spectrograph on board the AKARI and Spitzer space telescopes, respectively. 
To analyze the ice absorption features in the 8$-$\micron{22}, we first performed a continuum determination process on mid-infrared spectra and applied a method to subtract the silicate absorption.
We conducted a global fitting process to the absorption features in the combined infrared spectra using the experimental ice absorbance data to identify the intrinsic absorption of each ice component.
We first derived the \ch{H2O} ice column densities of both stretch and libration modes at \micron{3.05} and \micron{13.6} simultaneously.
We also identified the absorption features containing \ch{NH3}, \ch{CH3OH}, \ch{CO2}, and \ch{CO} and decomposed their mixed components and compared their ice abundances at different evolutionary stages of the protostars.
We explored possible absorptions of the organic ice species such as \ch{HCOOH}, \ch{CH3CHO}, and \ch{CH3CH2OH} in the mid-infrared ranges. The ice analysis method developed in this study can be applied to the ice spectra obtained by the James Webb Space Telescope.
}

\begin{document}
\jkashead 


\section{Introduction}
Ices in interstellar environments have become essential in our understanding of star formation, as they are key ingredients of planet formation, which is the end product of star formation.
Ices can form either by the freeze-out of gaseous molecules onto grain surfaces \citep{2003ApJ...583..789L,2004ApJ...617..360L} in the prestellar and protostellar stages. 
The surface reactions can form even more complex molecules that serve as precursors to organic compounds \citep{2006A&A...457..927G,2008ApJ...682..283G,2020ApJS..249...26J}.
In addition, the study of interstellar ice species has provided valuable insights into the chemical and physical processes that occur in the cold and dense environments of the interstellar medium. 
In warmer regions, such as protostellar disks and hot cores, the ice mantles can evaporate and release their trapped molecules into the gas phase, where they can participate in chemical reactions and contribute to the formation of larger and more complex organic molecules \citep{1974IAUS...60..155G,1999A&A...351.1066D,2006A&A...457..927G,2008ApJ...682..283G,2010ApJ...718..832O}.
Therefore, detecting and characterizing ice components in protostellar environments can shed light on the chemical processes involved in forming habitable planets and the emergence of life.

Infrared spectroscopy is a powerful tool for analyzing the chemical composition of ice species, as it provides detailed information on the absorption features of various molecules. 
Infrared space telescopes, including AKARI \citep{2007PASJ...59S.369M} and Spitzer \citep{2004ApJS..154....1W}, have been extensively used for studying interstellar ice through spectroscopic observations.
The 2.5$-$\micron{5.0} near-infrared spectra obtained by the Infrared Camera (IRC) on board the AKARI grism mode \citep{2007PASJ...59S.401O,2007PASJ...59S.411O} show significant absorption features toward low-mass protostars and several dense cores, which allow us to study the ice components and abundances, especially the \ch{CO} ice, which only shows the absorption at \micron{4.67} \citep{2012A&A...538A..57A,2013ApJ...775...85N,2022ApJ...935..137K}.
For the ice absorption features at longer infrared ranges, the 5$-$\micron{35} mid-infrared spectra observed with the Infrared Spectrometer \citep[IRS;][]{2004ApJS..154...18H} at the Spitzer have identified various ice species enabling to study of the formation of organic molecules in various environmental conditions \citep{2008ApJ...678..985B,2008ApJ...678.1032O,2010ApJ...718.1100B,2011ApJ...740..109O}.
The successful launch of the James Webb Space Telescope (JWST) and recently reported observations with high sensitivity and spectral resolution further expand the limits of the observational approach to ice chemistry through infrared spectroscopy \citep{2022ApJ...941L..13Y,2023A&A...673A.121B,2023NatAs...7..431M}.

Spectroscopic studies using continuous infrared spectra are particularly important for identifying and characterizing ice species in their various vibrational modes, which produce different shapes, strengths, and profiles of absorption features at each corresponding wavelength. 
Using the Infrared Space Observatory (ISO) survey, \citet{2004ApJS..151...35G} constrained the abundance of ice components such as \ch{H2O}, \ch{CO2}, and \ch{CH3OH} in various wavelengths through a comprehensive ice analysis that considered silicate components as well as vibration modes of various ice components using the 2.4 to \micron{45} spectra towards high-mass protostars.
However, the observed targets in the ISO survey were insufficient to examine the chemical evolution of ice components and their changes experienced during the low-mass star-forming history, despite the higher spectral resolution compared to other space telescopes.

This paper presents the combined AKARI and Spitzer spectra between 2.5 and \micron{30}, containing the absorption features of ice species in the circumstellar envelopes around four low-mass protostars at different evolutionary stages.
Section 2 describes the spectroscopic data acquisition of our targets for both AKARI IRC and Spitzer IRS.
Section 3 explains the continuum fitting process in the mid-infrared range, where we subtract silicate dust absorption to extract the ice absorption features from the Spitzer spectra.
This section also presents the extracted infrared spectra between 2.5$-$\micron{22} to identify absorption features of prominent or possible ice components.
Section 4 examines the ice composition of the absorption features based on a water-rich ice mixture using experimental data and estimates the ice column densities by fitting these ice absorbance profiles.
Section 5 discusses the abundance variation of some ice species, which are sensitive to the thermal conditions in the protostellar envelope and influenced by episodic mass accretion processes.

\section{Observations}
We obtained both near- and mid-infrared spectra for four protostellar targets (Per-emb 25, Ced 110 IRS4, B1 a, and RNO 91) to conduct the ice analysis of the absorption features from the infrared spectral ranges. 
The absorption features of the most abundant ice species (\ch{H2O}, \ch{CO2}, and \ch{CO}) can be observed by the AKARI IRC in the near-infrared range (2.5$-$\micron{5.0}) within the ${1}^\prime\times{1}^\prime$ field of view \citep[${1.}^{\prime\prime}$46 per pixel;][]{2007PASJ...59S.411O}, with a resolving power of R $\sim$ 120 at \micron{3.6} \citep{2013ApJ...775...85N}.
In the mid-infrared range (5.0$-$\micron{30}), the Spitzer IRS observed the absorption features of the \ch{H2O} and \ch{CO2} ices in different vibration modes \citep{2008ApJ...678..985B,2008ApJ...678.1005P}. 
The \ch{NH3} umbrella and \ch{CH3OH} stretching modes have also been observed by the Spitzer IRS spectra \citep{2010ApJ...718.1100B}. 

\subsection{AKARI IRC spectra}\label{sec:2a}
\begin{figure*}[t]
\centering
\includegraphics[width=170mm]{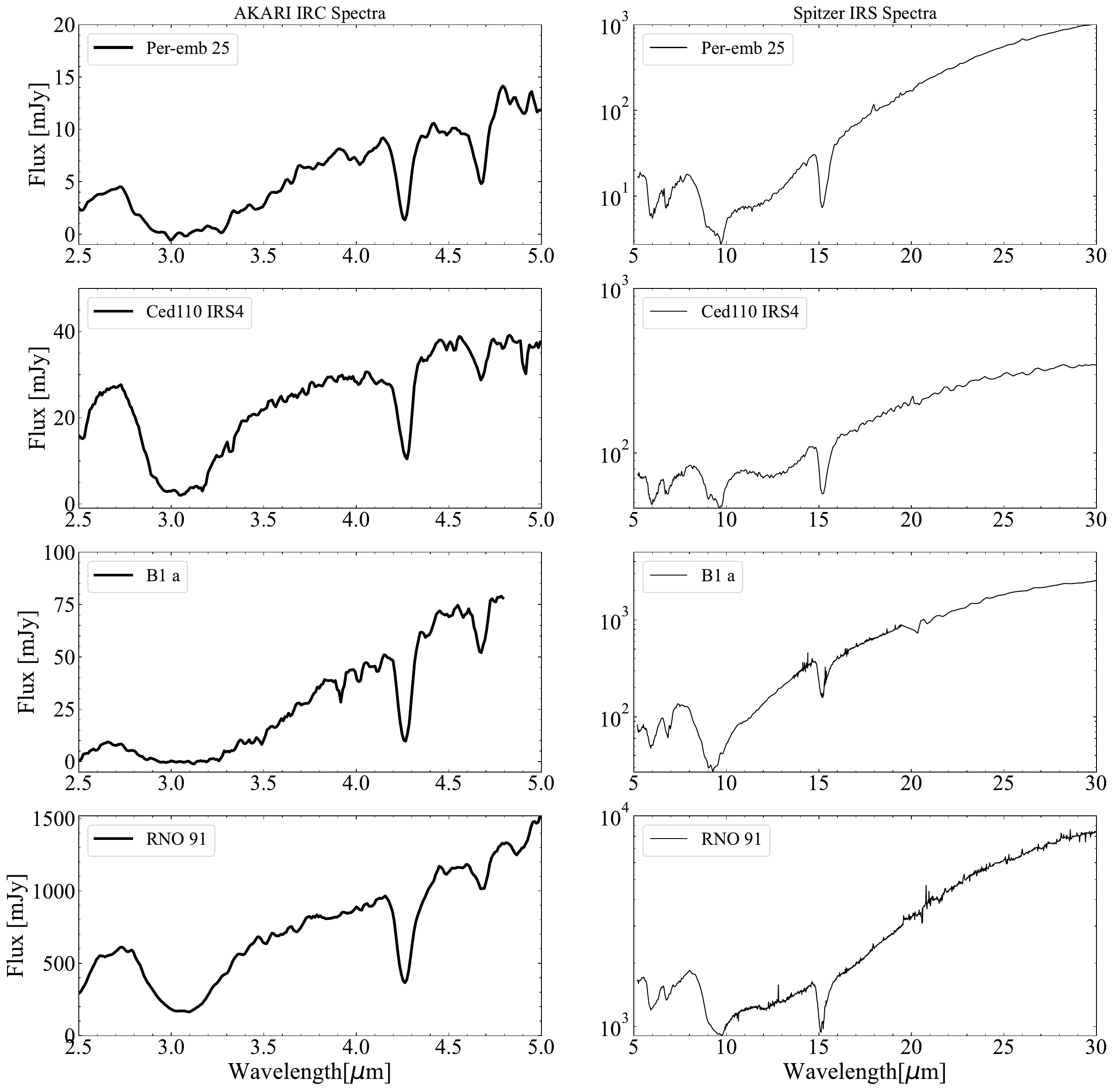}
\caption{The observed AKARI IRC spectra (left) and Spitzer IRS spectra (right) for all the sources in this study. In the case of RNO 91, we used the IRC spectrum reduced with short-exposure data to avoid the flux saturation of the target.
The Spitzer spectra are plotted in log scale while the AKARI spectra are plotted in linear scale.\label{fig:1}}
\vspace{5mm} 
\end{figure*}

\citet{2022ApJ...935..137K} carried out data reduction of the AKARI IRC spectroscopic data for Perseus 1 (called Per-emb 25 in this work) and RNO 91. 
For the spectra of Ced 110 IRS4 and B1 a, we followed the data reduction process of \citet{2022ApJ...935..137K} to obtain the reduced spectrum of those targets. 
We summarized their positions, evolutionary stages, observation IDs, and the corresponding spectroscopic modes in Table~\ref{tab:1}.
In the left panels of Figure~\ref{fig:1}, the reduced AKARI IRC spectra of our targets are presented, containing the absorption features for the stretching mode of the \ch{H2O}, \ch{CO2}, and \ch{CO} ices around \micron{3.05}, \micron{4.27}, and \micron{4.67}, respectively.
We exclude the data points at the center of the \ch{H2O} ice absorption from the ice analysis because half of our targets have almost zero flux levels around the \micron{3} absorption peak position.

\subsection{Spitzer IRS spectra}\label{sec:2b}
We obtained reduced Spitzer IRS spectra of our targets from the Spitzer IRS data archive in the IRSA database \citep{2003PASP..115..965E,2014Evans}.
The spectroscopic observations of the targets have been performed for the wavelength ranges of 5.0$-$\micron{10} and 10$-$\micron{30} through the Short-Low (SL) and Long-Low (LL) modules of Spitzer IRS, respectively, with the spectral resolving power of R $\sim$ 60$-$130, except for RNO 91.
We obtained higher quality of the reduced spectrum of RNO 91 with a resolving power of R $\sim$ 600 (LH), covering the 10$-$\micron{30} region. 
We also summarized their observation IDs and the corresponding modules in Table~\ref{tab:1}.
In the right panels of Figure~\ref{fig:1}, the target spectra show various absorption features, primarily due to ice, as well as the strong emission features from [\ch{Ne} \RomanNumeralCaps{2}] \citep[\micron{12.8}; RNO 91;][]{2010A&A...519A...3L} and [\ch{Fe} \RomanNumeralCaps{2}] \citep[\micron{17.9}; Per-emb 25 and RNO 91;][]{2010A&A...519A...3L}. 
They also contain strong \micron{9.7} and broad \micron{18} absorption features by silicate dust components.

\section{Results}
\subsection{Continuum Determination}
It is critical to determine the continuum corresponding to the spectral features to measure the absorption of ice species in the infrared spectra toward protostellar envelopes.
We determined the AKARI IRC spectra continuum using third- to fourth-order polynomials.
The fitting ranges consisted of four wavelength ranges: 2.65$-$2.75, 3.85$-$4.05, 4.4$-$4.6, and 4.8$-$\micron{4.95}.
We selected continuum points from those fitting ranges to avoid absorption features, varying slightly for each object to optimize the continuum levels of each spectrum.

Unlike the near-infrared spectrum, determining the continuum for the mid-infrared spectrum requires a distinct approach due to the limited availability of wavelength ranges that can reliably constrain the spectral continuum without absorption by the extinction of dust or ice.
Additionally, the absorption features observed beyond \micron{8} are often blended with the broad absorption band caused by silicate dust, further complicating the estimation of a reliable baseline.
To address this, we employed a method to subtract the absorption by the silicate dust from the observed mid-infrared spectra within the wavelength range of 8 to \micron{22} before we analyzed the absorption features only by the molecular ice components.

We adopted the spectrum of GCS 3 \citep{2004ApJ...609..826K} as the silicate template because it has been used as the best-known spectrum representing silicate dust to determine the absorption features of ice species from the Spitzer IRS spectra \citep{2008ApJ...678..985B,2009ApJ...694..459Z,2010ApJ...718.1100B}.
However, in some cases, the GCS 3 template did not sufficiently reflect the silicate absorption between 9 and \micron{10} \citep{2008ApJ...678..985B,2011A&A...526A.152V}.
A synthetic silicate spectrum composed of amorphous pyroxene and olivine profiles has been used to fit the silicate absorption to avoid the uncertainties of the silicate template \citep{2011ApJ...729...92B,2015ApJ...801..110P,2023NatAs...7..431M}.
Therefore, we used the silicate absorption profile combined with those amorphous components to subtract the silicate residual between 8 and \micron{10}.
The absorption spectra of amorphous olivine (MgFeSiO$_4$) and pyroxene (Mg$_{0.7}$Fe$_{0.3}$SiO$_3$) are synthesized using \texttt{optool} \citep{2021Dominik} and the lab measurements from \citep{1995A&A...300..503D}.  
We calculate the dust absorption using a $a^{-3.5}$ power-law grain size distribution with a minimum and maximum grain size of 0.1 and 1 $\mu$m, respectively.  
We also assume an 87\%\ of silicate dust and 13\%\ of carbon grains by mass.
The combined synthetic silicate spectrum was fitted by matching the spectral absorption features around \micron{9.7} and \micron{18}.

\begin{figure*}[t]
\centering
\includegraphics[scale=0.5]{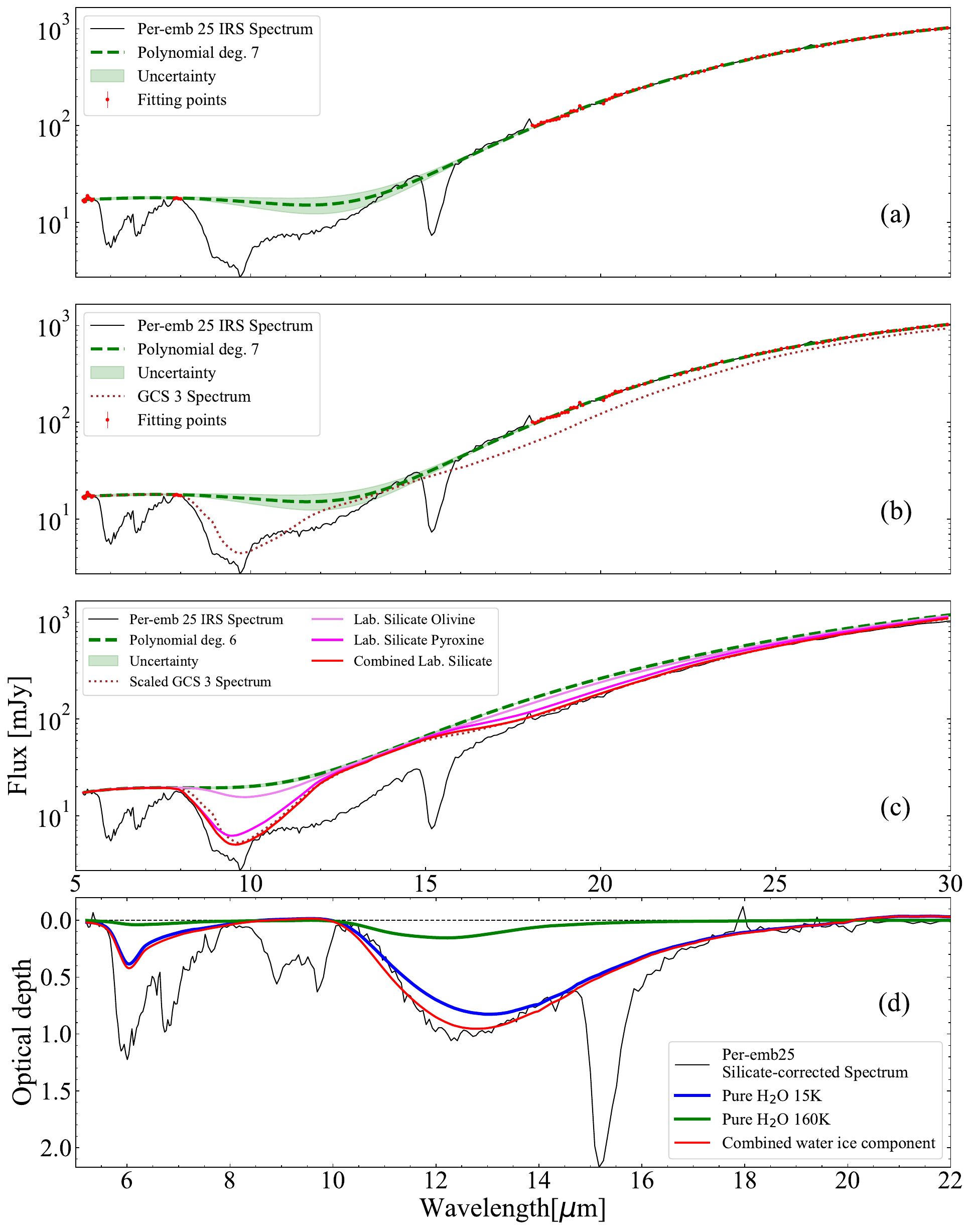}
\caption{An example of the continuum determination process for Per-emb 25. Panel (a) presents the initial continuum fit to the target spectrum using a six-order polynomial (green dashed line) with the selected wavelengths marked by red points. In Panel (b), the GCS 3 template spectrum (brown dotted line) is overlaid onto the fitted continuum to examine the absorption of silicate features to the target spectrum. Panel (c) shows the optimized continuum obtained by scaling the silicate absorption (red line) to match the observed spectral features, using a combination of pyroxene (violet line) and olivine (pink line) components. The green-shaded region indicates the estimated uncertainty in the optimized continuum, demonstrating that the continuum determination yields a robust result for the mid-infrared spectrum. Panel (d) displays the optical depth spectrum after subtraction of the silicate absorption features, thereby revealing the underlying ice absorption profile. The best-fit result of the \ch{H2O} ice absorption in libration mode is described as the red line. The fitted laboratory profiles well cover the broad \micron{13} absorption with the combined pure amorphous (blue line) and crystalline (green line) components.\label{fig:2}}
\vspace{5mm} 
\end{figure*}

The continuum determination process for the Spitzer IRS spectrum of Per-emb 25 is presented in Figure~\ref{fig:2} as an example, with a detailed description as follows. 
First, we selected continuum points at 5.2$-$5.5, 7.8$-$8.0, and 18$-$\micron{30} to interpolate a six-order polynomial function as an initial continuum, shown as a green dashed curve in Figure~\ref{fig:2}a.
Next, we placed the spectrum of GCS 3, used as the initial template for the silicate absorption, on the initial continuum (Figure~\ref{fig:2}b).
We then performed the interpolation process again by fitting the scaled silicate template to the observed spectrum.
In Figure~\ref{fig:2}c, a synthetic silicate spectrum created by combining amorphous pyroxene and olivine profiles was applied to correct the absorption by silicate components in the mid-infrared range and subtracted from the absorption features.
Finally, we used the experimental ice absorbance data from \citet{2004AdSpR..33...14F} to fit the silicate-subtracted absorption feature, focusing on the \ch{H2O} libration mode spanning 10$-$\micron{20}, using a mixture of amorphous and crystalline pure \ch{H2O} ices at 15 K and 160 K, respectively.

\begin{figure*}[t]
\centering
\includegraphics[scale=0.5]{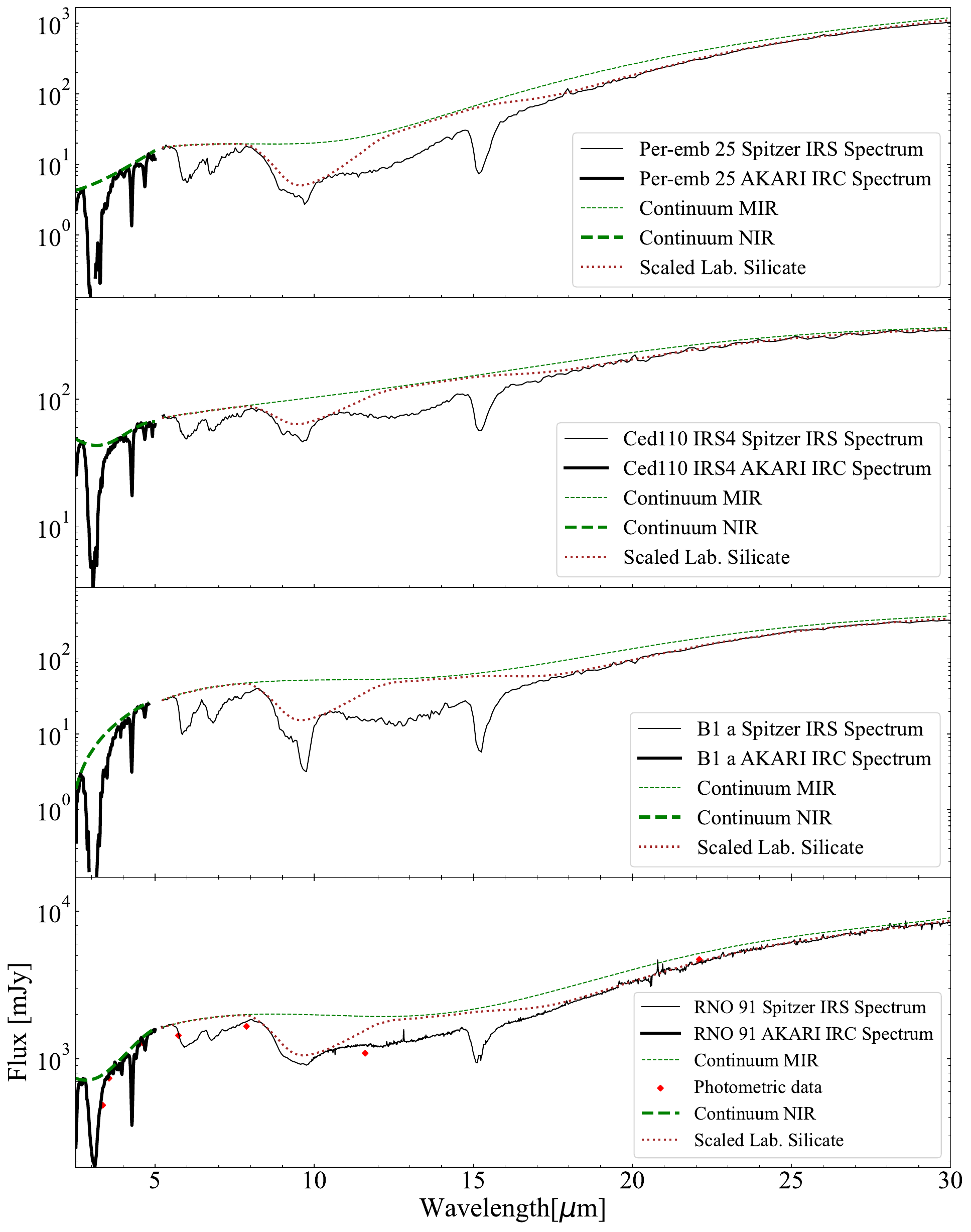}
\caption{
Combined infrared spectra from the AKARI IRC and Spitzer IRS observations for all the sources in this study. The thick green dashed line shows the continuum determined by third- to fourth-order polynomial fitting of the AKARI spectrum. Since the flux levels of the AKARI and Spitzer spectra are not identical, we scaled the AKARI spectrum to match the flux level of the short-wavelength end of the Spitzer spectrum.
The silicate-corrected continuum determined by sixth-order polynomial fitting of the Spitzer IRS spectrum is described as the thin green dashed line, overlaid with the scaled silicate absorption of the combined synthetic profiles (brown dotted line). For RNO 91, photometric data toward the target are plotted as red diamonds \citep{2021ApJS..253....8M,2021SEIP}. 
\label{fig:3}}
\vspace{5mm} 
\end{figure*}

The uncertainty inherent in the continuum fitted along with the silicate absorption components can affect the column density measurements of \ch{H2O} ice in the libration mode. 
However, the wavelength range where the silicate absorption components are mainly distributed is outside the region where the absorption features of \ch{H2O} ice appear. 
Beyond \micron{18}, most absorption features are predominantly due to silicate dust \citep{2010ARA&A..48...21H}. 
Therefore, we reduced the uncertainty in continuum determination by minimizing the difference between the spectrum and silicate absorption in this wavelength range. 
Additionally, we confirmed that the scaled silicate absorption feature around \micron{8} also matched well with the spectrum. 
To further improve the accuracy of the continuum determination, we included the 5.2 to \micron{5.5} range, which shows minimal dust and ice absorption features, in the fitting points.
The calculated column densities of the \ch{H2O} ice in Table~\ref{tab:3} indicate that the fitted \ch{H2O} absorption at the libration mode has relatively low uncertainty compared to its stretching mode.
Figure~\ref{fig:2}d shows that the fitted profile of the pure \ch{H2O} ice mixture well covered the broad absorption feature of the \ch{H2O} libration mode.
The optimized continuum of the combined AKARI and Spitzer spectra and the fitted silicate absorption for each target are overlaid as dashed and dotted curves in Figure~\ref{fig:3}, respectively.
Since the flux levels of the AKARI and Spitzer spectra are not identical, we scaled the AKARI spectrum to match the flux level of the short-wavelength end of the Spitzer spectrum. 

For RNO 91, a short-exposure AKARI spectrum was used to avoid flux saturation around the \micron{3} absorption feature.
However, the resulting flux level was lower than expected, so we scaled the spectrum upward based on the corresponding long-exposure data to ensure a more reliable spectral analysis. 
To validate this adjustment, we plotted the photometric measurements in the bottom panel of Figure~\ref{fig:3}, which show good agreement with the scaled spectrum, supporting the appropriateness of the applied correction.
In the case of the silicate absorption removal, no significant difference was found in the 9 to \micron{10} range when using the synthetic silicate spectrum compared to using the GCS 3 silicate template. However, the GCS 3 template provided a better fit to the \ch{H2O} ice absorption feature at the libration mode. 
Therefore, for RNO 91, we applied the silicate subtraction by scaling the GCS 3 silicate template, unlike for the other three targets.

\subsection{The Ice Absorption Features}\label{sec:3b}
\begin{figure*}[t]
\centering
\includegraphics[scale=0.5]{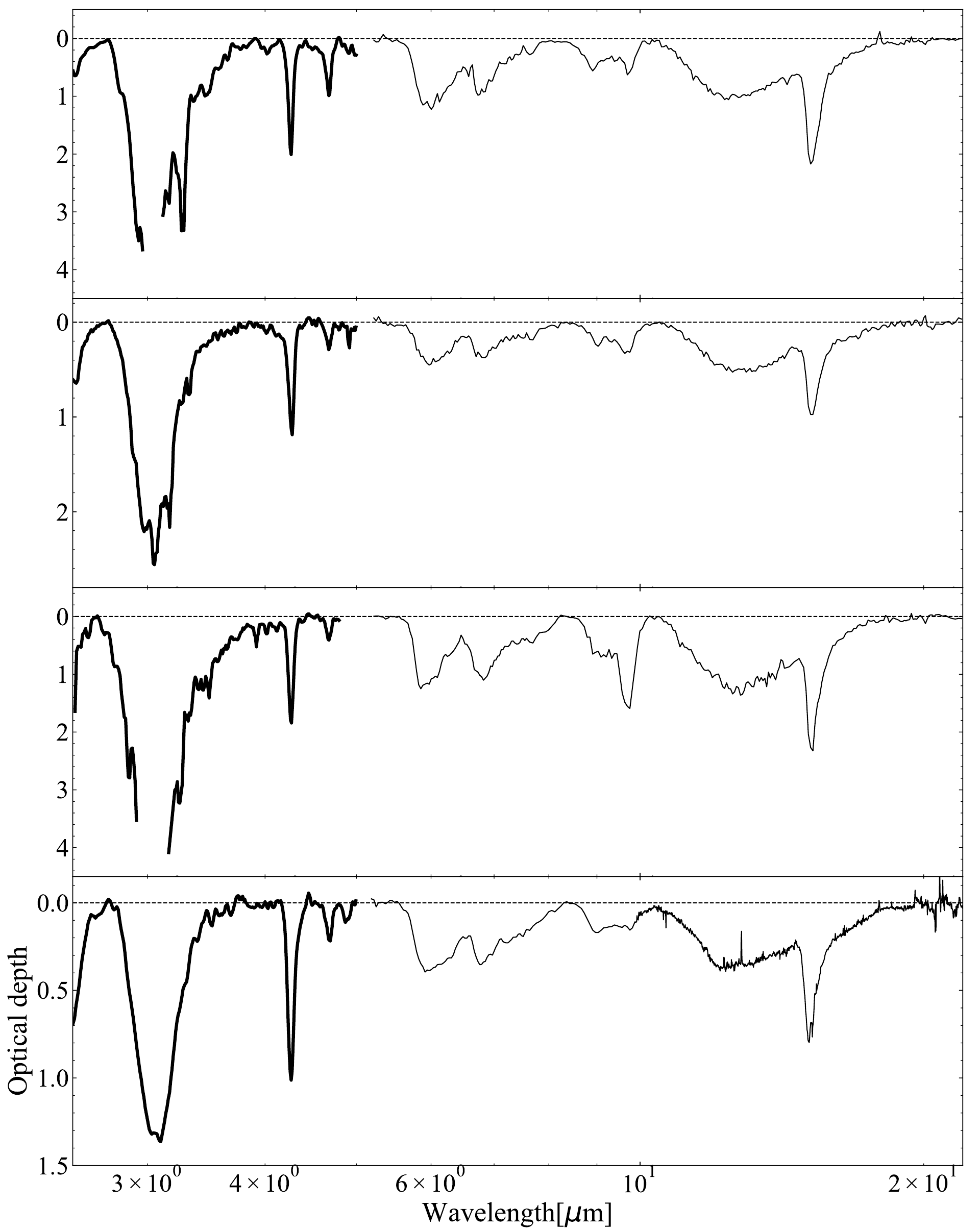}
\caption{Same spectra as Figure~\ref{fig:3}, but for the corresponding optical depth plot calculated from the equation, $\tau$ $=$ ln($I_{0}$/$I$), where $I_{0}$ and $I$ are the continuum and observed fluxes, respectively. 
\label{fig:4}}
\vspace{5mm} 
\end{figure*}

Figure~\ref{fig:4} presents the silicate-subtracted spectra of the four targets on an optical depth scale, revealing the absorption features of various ice species.
Our targets exhibit distinct and broad absorption features corresponding to the stretching and libration modes of \ch{H2O} ice in the wavelength ranges of 2.7$-$\micron{3.4} and 10$-$\micron{20}, respectively. 
The bending mode of \ch{H2O} ice at \micron{6.0} is blended with organic ice components like \ch{HCOOH}, \ch{H2CO}, and \ch{CH3CHO}, resulting in complex absorption features \citep{2011ApJ...729...92B,2011ApJ...740..109O,2019A&A...626A.118Q}.

The bending mode of \ch{CO2} ice around \micron{15} showed deep absorption with a shoulder extending towards longer wavelengths.
The stretching mode at \micron{4.27} shows a narrower and deeper absorption feature compared to the bending mode.
Previous ice studies employing the AKARI IRC have addressed the challenges of unresolved absorption features using various techniques.
\citet{2010A&A...514A..12S} utilized the curve-of-growth method to estimate the column density of the narrow absorption features, while others convolved experimental ice profiles to match the instrumental resolution of the AKARI IRC \citep{2012A&A...538A..57A,2013ApJ...775...85N,2021ApJ...916...75O,2022ApJ...935..137K}.
In our study, we adopted a similar convolution approach based on experimental ice profiles derived from the ice analysis process described by \citet{2022ApJ...935..137K}.
However, our study benefits from a well-resolved bending mode of \ch{CO2} ice, which enables a more robust estimation of the \ch{CO2} ice composition \citep{2008ApJ...678.1005P}.
Moreover, as recent JWST observations of Ced 110 IRS4 revealed that the \ch{CO2} stretching mode exhibits an exceptionally deep absorption feature \citep{2025A&A...693A.288R}, our findings underscore the advantage of analyzing spectra with the bending mode absorption rather than relying on the unresolved stretching mode features in AKARI IRC.

In the infrared range, the absorption feature of CO ice is solely observed in the stretching mode around \micron{4.67}.
The absorption feature of CO ice in protostellar envelopes generally shows a broad wing structure, which overlaps with XCN ice and gaseous CO components \citep{2012A&A...538A..57A,2021ApJ...916...75O,2022ApJ...935..137K}.
The \micron{6.85} absorption feature, observed in all targets alongside the \micron{6.0} absorption, is attributed to \ch{CH3OH} ice \citep{2004ApJS..151...35G,2008ApJ...678..985B,2022ApJ...930....2M,2022ApJ...941L..13Y,2023NatAs...7..431M}.
Another potential absorber, \ch{NH4+}, may contribute to the absorption feature.
It is suggested that this component is likely produced in environments affected by energetic processing with cosmic rays and heating from the central protostar \citep{2003A&A...398.1049S,2008ApJ...678..985B,2011ApJ...740..109O}.
In the 8$-$\micron{10} range, where most of the absorption by the silicate components has been subtracted from the mid-infrared spectrum, broad absorption features are detected, primarily attributed to the ice components of the \ch{NH3} umbrella mode and the \ch{CH3OH} stretching mode \citep{2010ApJ...718.1100B}.

\section{Comprehensive Ice Analysis}
Once an optical depth spectrum was derived, we conducted a fitting process using the experimental ice absorbance data \citep{1995A&A...296..810G,1996A&A...315L.341E,2004AdSpR..33...14F} convolved with the spectral resolutions of observations to identify the intrinsic absorption profiles of each ice component that composes the absorption features.
However, due to ice composition variations and blended features, it is important to determine which ice component takes precedence.
We first proceeded with fitting the \ch{H2O} ice component which has the most significant abundance among the ice species, considering the previous studies for the ice abundance distribution of the embedded protostars \citep{2008ApJ...678..985B,2015ARA&A..53..541B}.
Derivation of the intrinsic absorption profiles for other abundant ice species proceeded to the ice components of \ch{CO2}, \ch{CO}, \ch{CH3OH}, and \ch{NH3} in a sequential manner.
Additionally, we accounted for minor constituents, including \ch{NH4+}, \ch{HCOOH}, \ch{CH3CHO}, and \ch{CH3CH2OH} ices, which might be mixed with the bending mode of \ch{H2O} ice and other ice absorptions.
Table~\ref{tab:2} shows the references of experimental ice absorbance profiles used for the ice composition of the absorption features.

\subsection{\ch{H2O}}
\begin{figure*}[t]
\centering
\includegraphics[scale=0.5]{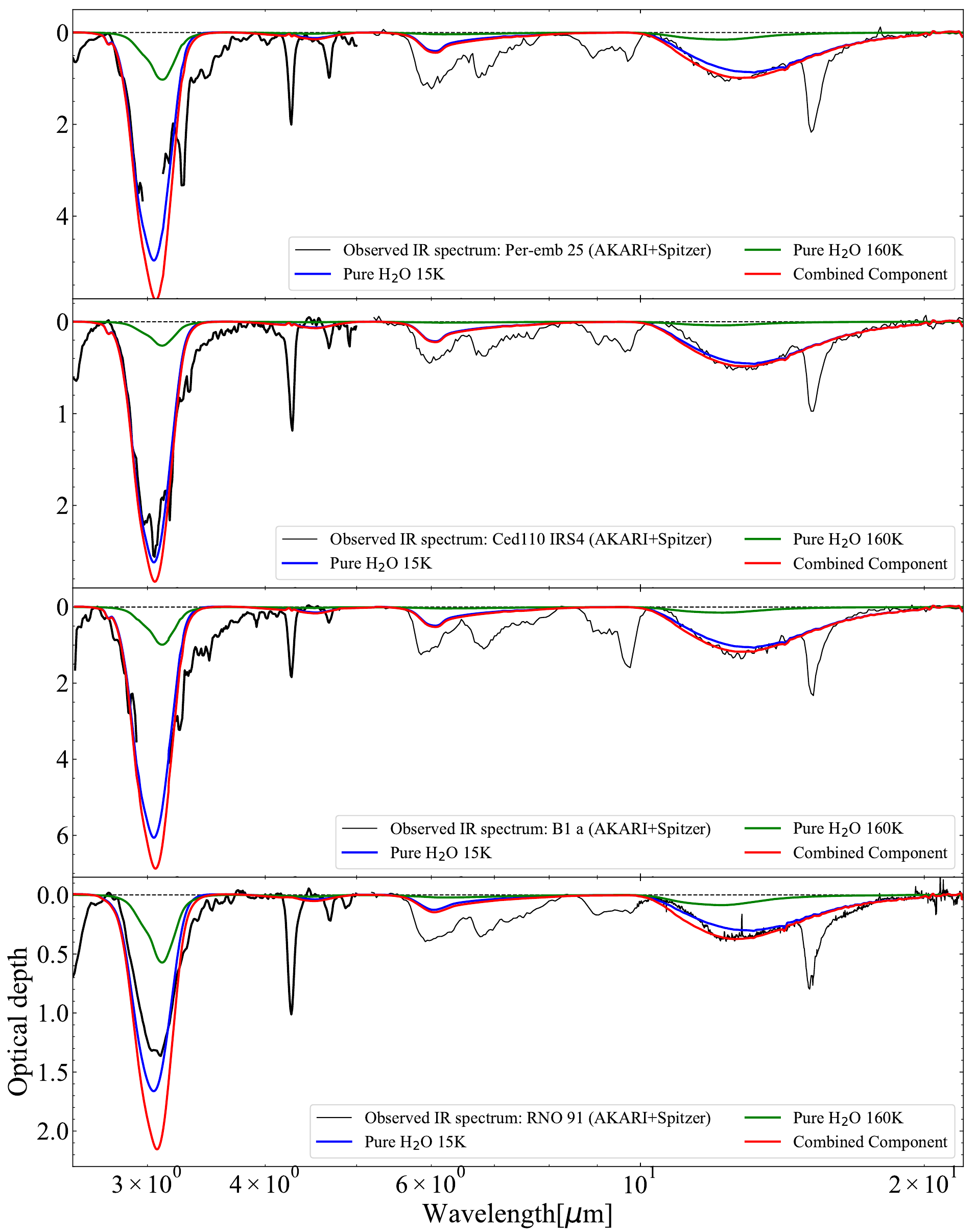}
\caption{Same spectra as Figure~\ref{fig:4}, but for the best-fit results of the \ch{H2O} ice absorption features. The fitted laboratory profiles are described in the bottom right of each panel. The combined ice profile with the pure amorphous (blue line) and crystalline (green line) components is described as the red line. The fitted profile well covers the \ch{H2O} ice absorption features, especially for both stretch (peaked at \micron{3.05}) and libration (peaked at \micron{13.6}) modes.  \label{fig:5}}
\vspace{5mm} 
\end{figure*}

Figure~\ref{fig:4} clearly shows the absorptions of \ch{H2O} ice at the stretching, bending, and libration modes, as discussed in Section~\ref{sec:3b}.
However, to evaluate the overall fit quality of the \ch{H2O} absorption features, we focused on the libration mode covering the wavelength range of 10$-$\micron{22}.
This range was chosen because the bending mode of \ch{H2O} ice is blended with other ice absorptions, and for half of our targets (Per-emb 25 and B1 a), the spectrum around the central range of the stretching mode could not be extracted as we described in Section~\ref{sec:2a}.
Additionally, potential absorption components from \ch{CH3OH}-mixed ice at the O-H stretching mode around \micron{3.24} \citep{2023NatAs...7..431M} could affect the accuracy of the \ch{H2O} ice column density, especially for its crystalline component.
We used the amorphous and crystalline pure ice profiles at temperatures of 15 K and 160 K, respectively.
The best-fit results, represented by red solid lines in Figure~\ref{fig:5}, effectively cover the areas of both stretching and libration modes. 
However, in the case of RNO 91, the best-fit result at the stretching mode was overestimated compared to its libration mode.
This discrepancy might be due to the incomplete measurement of the \ch{H2O} absorption feature in the RNO 91 spectrum, even though we have conducted the short-exposure data to avoid the spectrum reaching the flux saturation level \citep[0.6 Jy at \micron{3};][]{2007PASJ...59S.401O} of AKARI IRC.
Indeed, \citet{1999ApJ...517..883B} showed an optical depth spectrum of RNO 91 from the ground-based \micron{3} observations for protostars, which has a sufficient absorption of the \ch{H2O} ice comparable to our best-fit result.

Crystalline ice component has been used to fit the ice composition of \ch{H2O} absorption to account for the thermal processing of ice due to protostellar heating \citep{2015ARA&A..53..541B,2021ApJ...916...75O,2022ApJ...935..137K}.
The existence of the crystalline ice component might be the main contribution to the shifted peak of the absorption feature by the stretching mode at \micron{3.10} (Bottom panel of Figure~\ref{fig:5}).

In addition to the pure \ch{H2O} ice composition, we considered \ch{H2O}-rich ice mixtures to interpret the absorption features that could not be fully explained by the pure ice component.
For example, the \ch{H2O}-rich \ch{NH3}-containing ice causes a shift in the absorption profile of pure \ch{NH3} from \micron{9.3} to \micron{9.0} \citep{2010ApJ...718.1100B}.
Similar shifts are observed in the absorption profiles of \ch{CH3OH} and \ch{CO2} ices when mixed with other ice components, including the \ch{H2O}-rich component \citep{2008ApJ...678.1005P,2010ApJ...718.1100B}.
Thus, we modified the ice composition of \ch{H2O} absorption by combining the water-rich mixture profiles containing \ch{CO}, \ch{NH3}, \ch{CH3OH}, and \ch{CO2} ices.
This comprehensive ice fitting approach covered the overall absorption features, and the cyan-colored solid lines in Figure~\ref{fig:6}$-$~\ref{fig:9} represent the best-fit results for the water-rich ice mixture of \ch{H2O}:\ch{CO}:\ch{NH3}:\ch{CO2}:\ch{CH3OH} $=$ 100:20:20:14:10 at 10 K.

\subsection{Other ice species}
\begin{figure*}[t]
\centering
\includegraphics[scale=0.44]{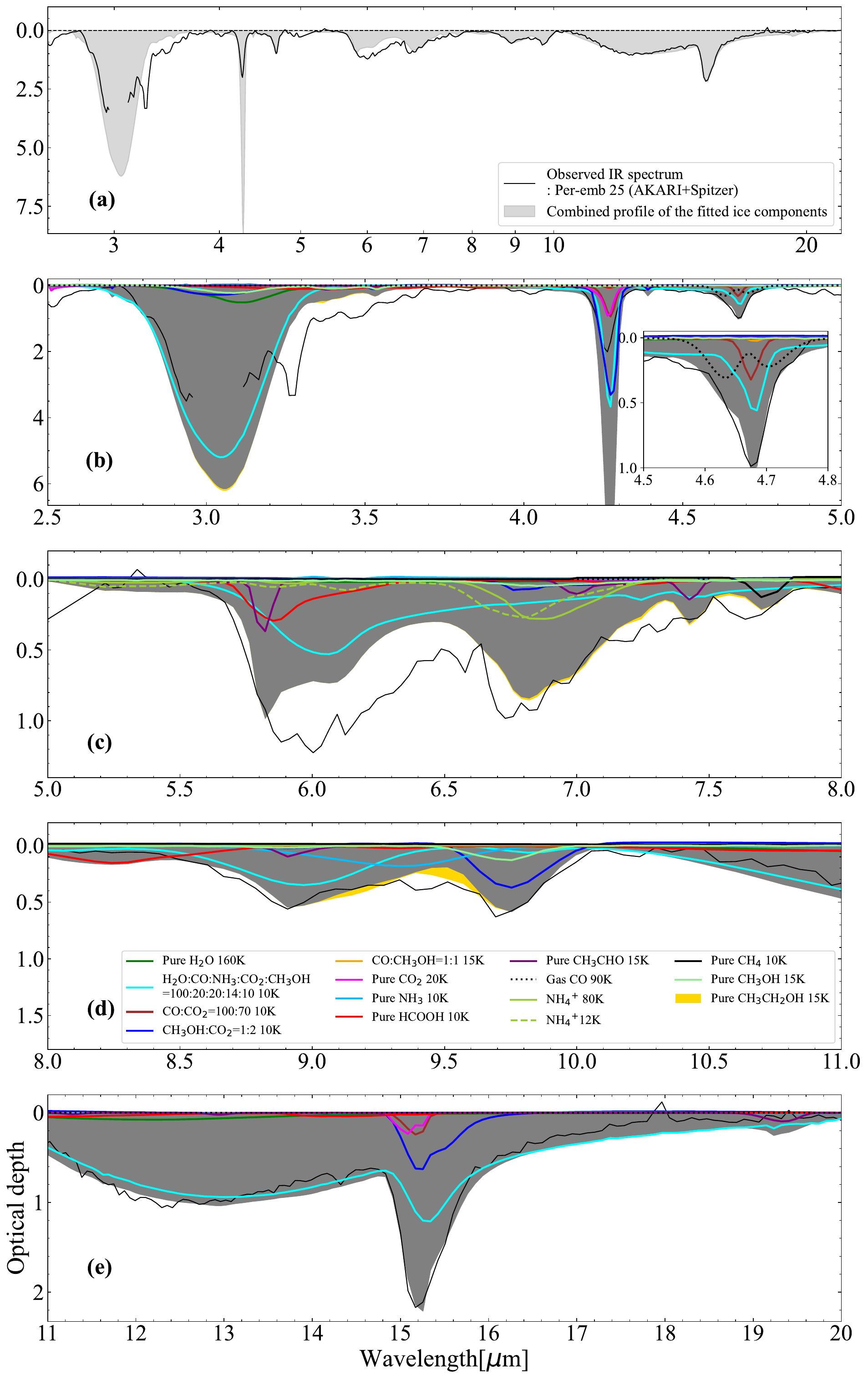}
\caption{(a) Same spectrum of Per-emb 25 in Figure~\ref{fig:4}, but for the best-fit results (gray-shaded region) of the combination of all ice components used in this study. The fitted laboratory profiles are described in panel (d). (b) Optical depth 2.5$-$\micron{5.0} AKARI IRC spectrum of the target shows the 3.05, 4.27, and \micron{4.67} absorption features of \ch{H2O}, \ch{CO2}, and \ch{CO} ices, respectively. (c$-$e) Optical depth 5.0$-$\micron{20.0} Spitzer IRS spectrum of the target shows 6.0, 9.0$-$9.75, 13.6, and \micron{15.2} absorption features of \ch{H2O} bend, \ch{NH3}$-$\ch{CH3OH} mixture, \ch{H2O} libration, and \ch{CO2} bend ices, respectively. \label{fig:6}}
\vspace{5mm} 
\end{figure*}

\begin{figure*}[t]
\centering
\includegraphics[scale=0.46]{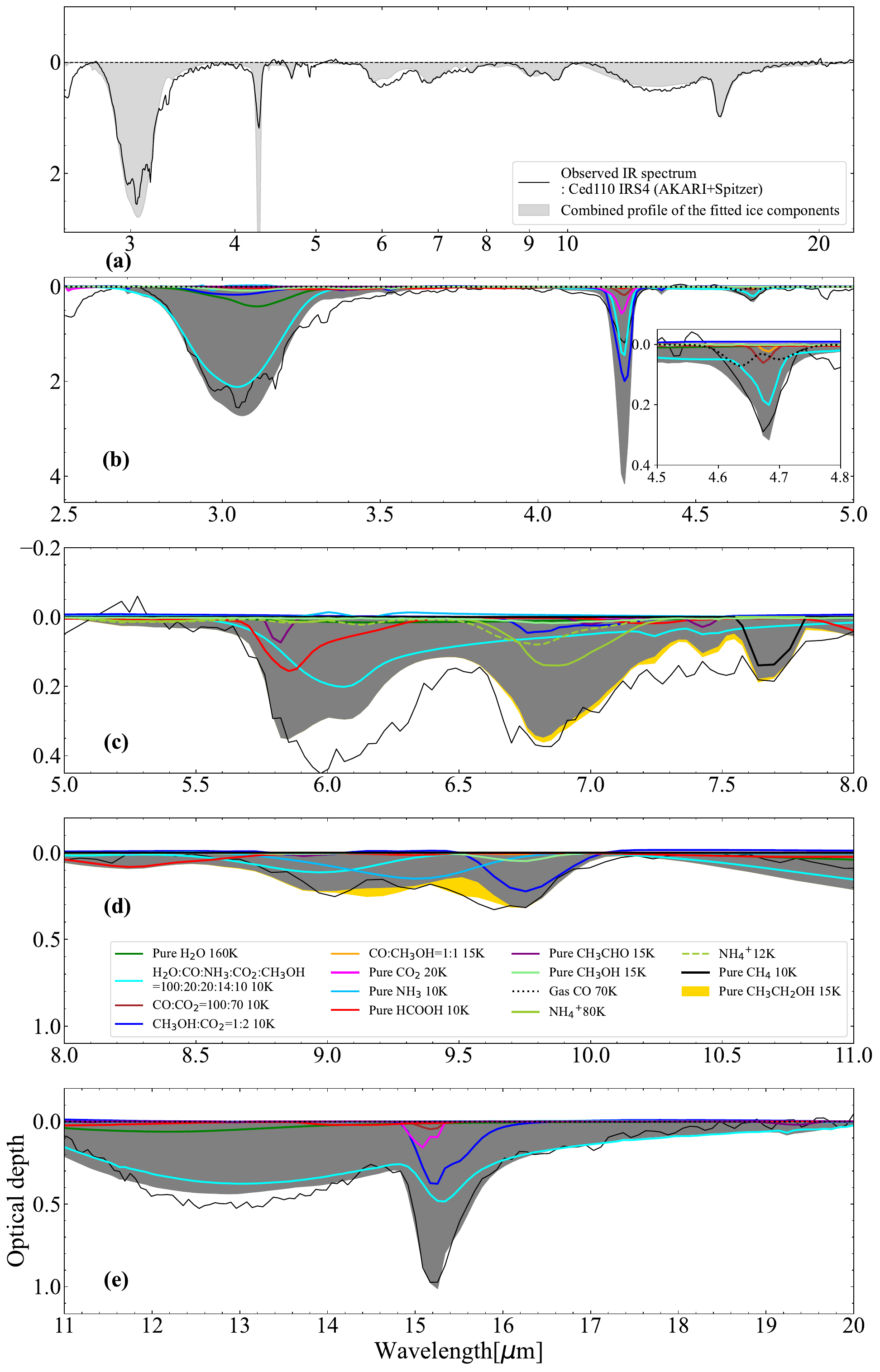}
\caption{The same plots as Figure~\ref{fig:6}, but for Ced 110 IRS4. \label{fig:7}}
\vspace{5mm} 
\end{figure*}

\begin{figure*}[t]
\centering
\includegraphics[scale=0.46]{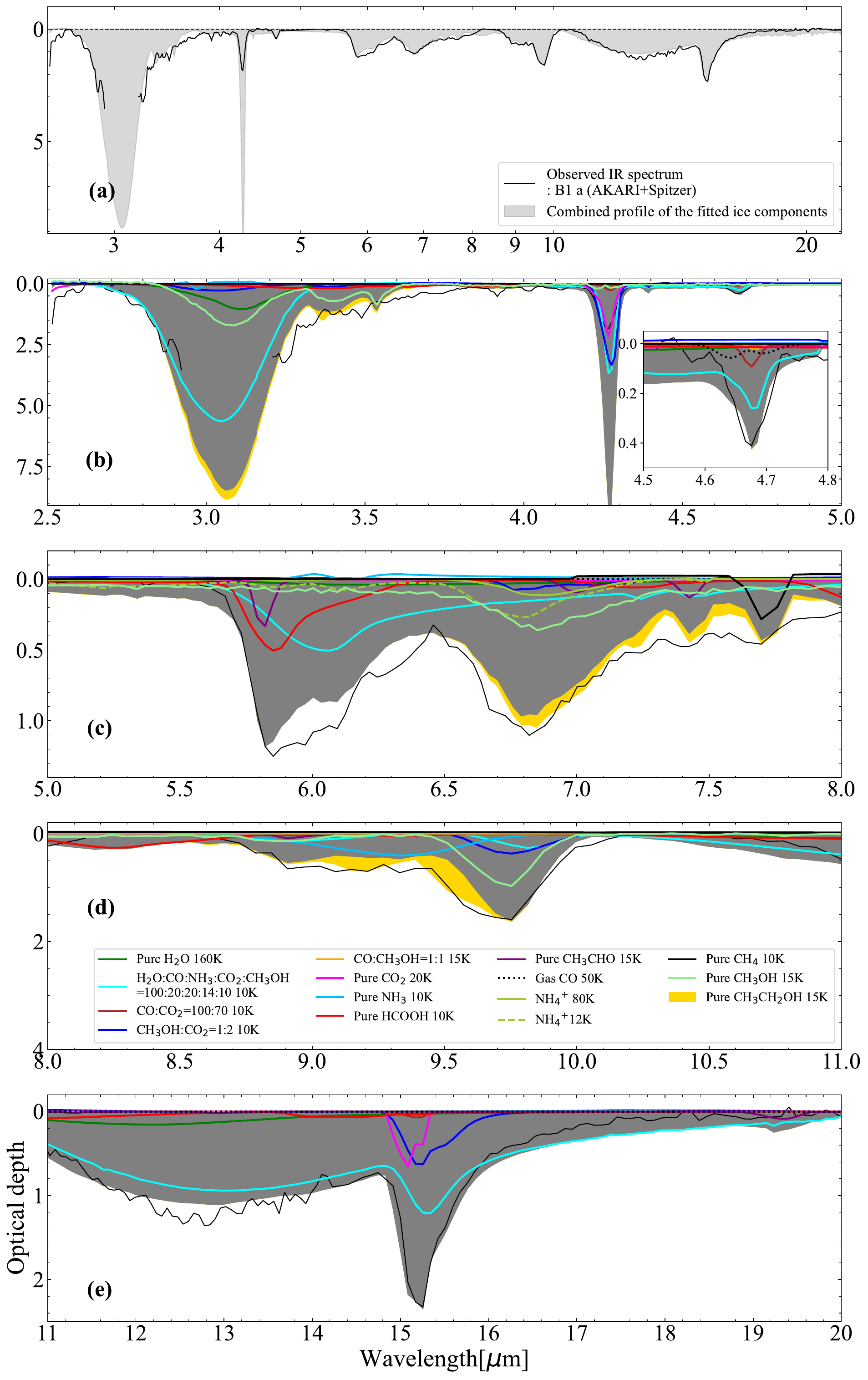}
\caption{The same plots as Figure~\ref{fig:6}, but for B1 a. \label{fig:8}}
\vspace{5mm} 
\end{figure*}

\begin{figure*}[t]
\centering
\includegraphics[scale=0.46]{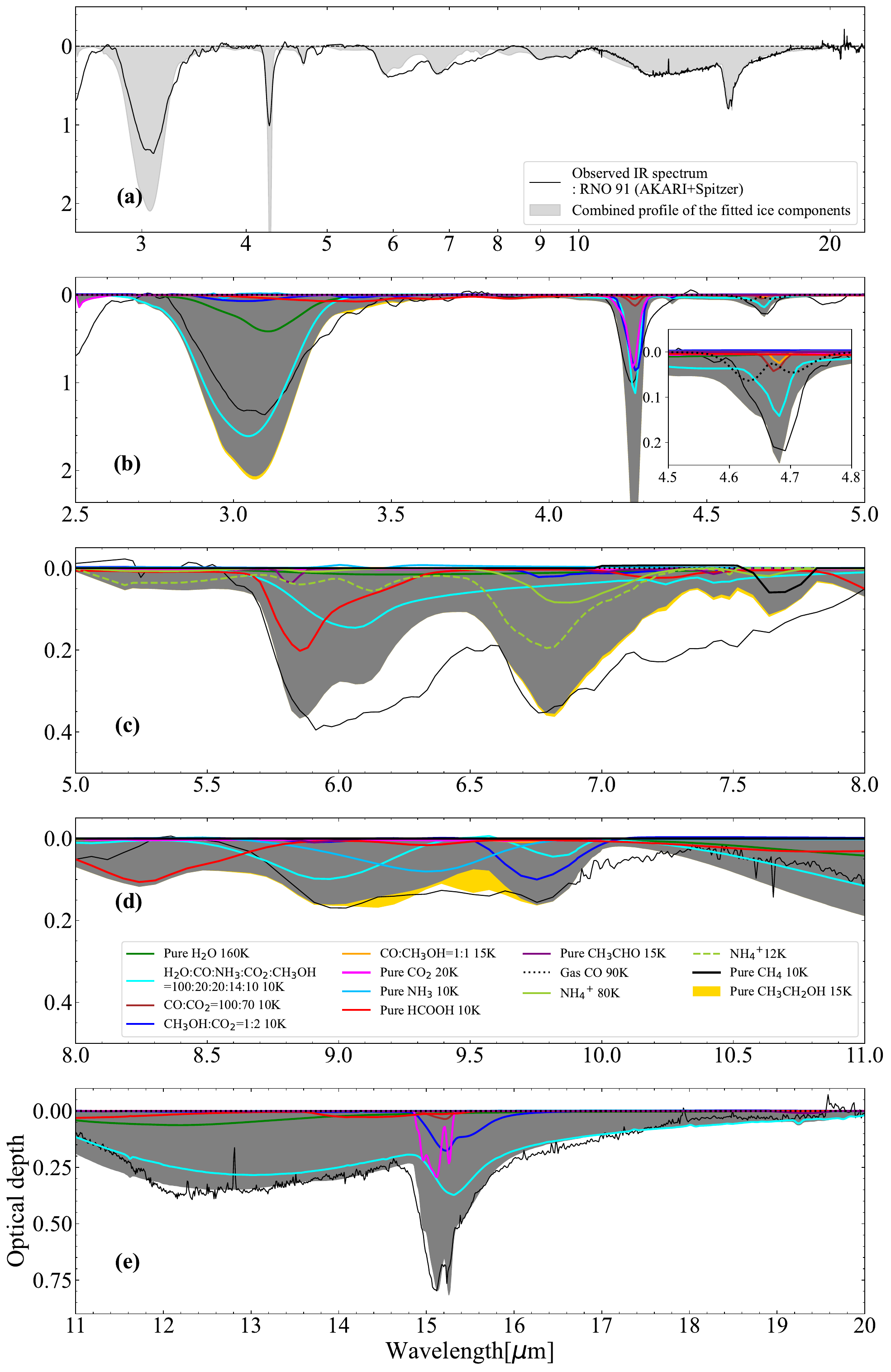}
\caption{The same plots as Figure~\ref{fig:6}, but for RNO 91. \label{fig:9}}
\vspace{5mm} 
\end{figure*}

We first adapted the \ch{CO}- and \ch{CH3OH}-mixed \ch{CO2} ice profiles (\ch{CO}:\ch{CO2} $=$ 100:70 at 10 K and \ch{CH3OH}:\ch{CO2} $=$ 1:2 at 10 K, respectively) to the \micron{15} absorption feature of the bending mode of \ch{CO2} along with the \ch{H2O}-rich ice component (cyan line in each bottom panel of Figure~\ref{fig:6}$-$~\ref{fig:9}).
These ice compositions have been suggested through the ice studies of chemical reaction pathways \citep{1997A&A...328..649E,1999A&A...351.1066D,2008ApJ...678.1005P}
Additionally, we employed pure \ch{CO2} ice, whose absorbance profile was corrected by using the continuously distributed ellipsoids grain model \citep{1999ApJ...522..357G}, to fit the double-peaked feature at \micron{15.2}, particularly in the high-resolution spectroscopic case of RNO 91.
Detection of the double-peaked feature has been suggested to be a chemical tracer of strong heating based on the thermal annealing process \citep{1999ApJ...522..357G,2018ApJ...869...41H}.
Pure \ch{CO2} ice might be left from the \ch{CO}-mixed layer after the thermal evaporation of \ch{CO} in the heated environment \citep{2012ApJ...758...38K}. 
Since this formation process is irreversible \citep{1983A&AS...51..389H}, the existence of pure \ch{CO2} ice component within the protostellar envelope could be indicative of past thermal history.
Recently, \citet{2021ApJ...919..116Y} reported that the broadened CO overtone absorption feature for the \textit{K} band spectra of RNO 91 is interpreted as evidence for an ongoing accretion burst. 
Therefore, we suggest the double-peaked \ch{CO2} ice feature can be another plausible evidence of the past and current accretion history of RNO 91.

For other targets, relevant analyses could not be performed due to the lack of spectral resolution.
The absorption features of \ch{CO} ice in our targets were generally fitted using a combination of \ch{H2O}-rich and \ch{CO2}-mixed ice profiles, along with the \ch{CO} gas absorption profile in the temperature range of 50 to 90 K (dotted line in the inset region of each panel (b) of Figure~\ref{fig:6}$-$~\ref{fig:9}).
The gaseous absorption feature of \ch{CO} has been reported for embedded protostars in other ice studies \citep{2012A&A...538A..57A,2021ApJ...916...75O,2022ApJ...935..137K}.

For the 8.5$-$\micron{10} absorption feature, we fitted \ch{NH3} and \ch{CH3OH} ice profiles with the applied \ch{H2O}-rich and \ch{CO2}-mixed components.
We also used pure \ch{NH3} ice and an \ch{CO}:\ch{CH3OH} mixture (1:1) to cover the absorption range between 9.0$-$\micron{9.7}.
We explored the possibility of the contribution of the \ch{CH3CH2OH} ice to the 9.0$-$\micron{9.8} wavelength range, as indicated by the yellow-shaded region in each panel (d) of Figure~\ref{fig:6}$-$~\ref{fig:9}.
The ice component of \ch{NH4+} was considered as the main contribution occupying the absorption around \micron{6.8}, although the identification of \ch{NH4+} is still subject to debate \citep{2003A&A...398.1049S,2022ApJ...941L..13Y,2023NatAs...7..431M}.
Finally, we employed pure \ch{HCOOH} and \ch{CH3CHO} ice profiles to compose the absorption feature around \micron{5.8}, where the bending mode of \ch{H2O} ice is blended.
Previous Spitzer and ISO observations \citep{1999A&A...343..966S,2008ApJ...678..985B}, as well as the recently reported JWST program \citep{2022ApJ...941L..13Y} have suggested the identification of those organic ice species from the spectral studies for the low- and high-mass protostars.

The gray-shaded regions in panel (a) of Figure~\ref{fig:6}$-$~\ref{fig:9} represent the best-fit results for these ice compositions across the entire 2.5$-$\micron{22} spectral range of each target spectrum.
Despite our comprehensive ice analysis of the infrared absorption features using a combination of pure and mixed ice profiles, a significant fraction of the 6$-$\micron{8} absorption remains unattributed, likely due to other organic and ion-based ice constituents.
Identifying these components with low resolving power is challenging, necessitating more powerful spectroscopic observations to confirm their contributions to the infrared absorption features.

\subsection{Ice abundances}
It is necessary to quantify the ice column density from the absorption features to understand the chemical condition of ice varying in the protostellar envelopes.
We derived the column density of each ice component for our targets using the fitted ice compositions.
The column density ($N$) of each ice component was obtained using the following equation,
\begin{equation}
 N = \frac{\int \tau d\nu}{A},\\\end{equation}
 where $A$ is the band strength (in cm molecules$^{-1}$; Table~\ref{tab:2}) and $\tau$ is the optical depth.
We integrated the optical depth of each ice component by combining the fitted experimental profiles in Table~\ref{tab:2}.

The column densities of the identified ice species derived from our analysis are summarized in Table~\ref{tab:3}.
We examined the \ch{H2O} ice column densities at both stretching and libration modes to validate which vibration mode is appropriate for calculating the ice abundances for other species.
At all targets, although the differences are within the uncertainties, the column density of the stretching mode of \ch{H2O} was higher than that of the latter. 
It might be interpreted as the absorption of other ice components mixed in the \ch{H2O} ice profile, such as the stretching mode of \ch{NH3} and \ch{CH3OH}, does not appear in the absorption region at the libration mode of \ch{H2O}.
We specify the vibration mode and mixture of each ice component used in the column density calculation in each column of Table~\ref{tab:3}.

\section{Discussions}
The relative ice abundances are calculated with respect to the \ch{H2O} ice because \ch{H2O} ice is the most abundant species formed before the protostellar stages.
Due to the possibility that the absorption feature around the \ch{H2O} stretching mode might be composed of other \ch{H2O}-rich ice components, such as \ch{CH3OH}, we used the column density of \ch{H2O} ice calculated in the libration mode rather than the stretching mode to derive the abundance of other ice species.
The derived ice abundances are given in Table~\ref{tab:4}.

We compared the results of Per-emb 25, B1 a, and RNO 91 with those previously reported for AKARI \citep{2022ApJ...935..137K} and Spitzer \citep{2008ApJ...678..985B,2008ApJ...678.1005P,2010ApJ...718.1100B,2012ApJ...758...38K} spectra to validate our comprehensive ice analysis.
Their abundances are also listed in Table~\ref{tab:4}.
The \ch{CO2} ice abundances calculated from the bending mode in our work are similar to those derived from the same vibration mode.
However, the result of Per-emb 25 is almost twice as high as the AKARI result obtained from the stretching mode.
A similar trend is observed for the \ch{H2O} ice column densities of Per-emb 25 and RNO 91 when compared to the previous AKARI study.
\citet{2022ApJ...935..137K} suggested that the underestimation of \ch{H2O} and \ch{CO2} ice column densities in the targets might be attributed to a saturation effect, which caused by the nearly zero flux level at the absorption peak, and in case of \ch{CO2}, to the low spectral resolution of the AKARI IRC.

It is insufficient to reveal clear trends linked to the differences in ice abundances across only the four samples and the evolutionary stages of protostars, even though they are in different stages of evolution.
However, the abundance of \ch{CO} ice tentatively decreases as the evolutionary sequence from Class 0 to \RomanNumeralCaps{1}. 
During the prestellar stages, as the freeze-out rate of gas-phase \ch{CO} molecules gradually increases with gas density, most \ch{CO} molecules freeze out on grain surfaces coated with \ch{H2O} ices, resulting in \ch{H2O}-rich \ch{CO} and \ch{CO}-rich \ch{CO2} ice mixtures \citep{2011ApJ...740..109O}.
As the protostar evolves and undergoes mass accretion, the heating of grain surfaces causes the sublimation of \ch{CO} ice due to its low sublimation temperature ($\sim$30 K).

\citet{2023NatAs...7..431M} recently reported the ice study of two dense regions (NIR38 $=$ 11:06:25.39, -77:23:15.70; J110621 $=$ 11:06:21.70, -77:23:53.50) of the Chameleon \RomanNumeralCaps{1} cloud that detect deep absorption features of \ch{H2O}, \ch{CO2}, and \ch{CO} ices using JWST spectroscopic data obtained with the Near Infrared Spectrograph \citep[NIRSpec,][]{2022A&A...661A..80J} Fixed Slit (FS) mode ($R \sim$ 2,600, 2.7$-$\micron{5.3}), as well as absorption features of organic ices (\ch{CH4}, \ch{NH4+}, \ch{NH3}, and \ch{CH3OH}) obtained with the Mid-Infrared Instrument \citep[MIRI,][]{2015PASP..127..584R} Low-Resolution Spectrograph (LRS) FS mode ($R \sim$ 100, 5$-$\micron{14}).
These regions provide a valuable opportunity to compare the abundance variation of ice inventory before and after the onset of star formation due to their proximity ($\sim{80}^{\prime\prime}$ from NIR38 and $\sim{110}^{\prime\prime}$ from J110621) to one of our Class \RomanNumeralCaps{1} targets, Ced 110 IRS 4.
The measured \ch{CO} ice abundances in these two dense regions were higher than those in our target, suggesting that the initially formed \ch{CO} ice components gradually converted to the gas phase in the heated protostellar inner envelope as the system evolved \citep{2004ApJ...617..360L}.
This observation implies a decrease in \ch{CO} ice abundance after the onset of the protostellar evolution.
In contrast, the abundance of the other ice species increased to the protostellar stage, which can be attributed to thermal processing and UV irradiation of ice mixtures during the protostellar phase.

It has been suggested that episodic accretion of surrounding material onto the central protostar leads to thermal ice processing and the formation of pure ice components from the ice mixtures, such as pure \ch{CO2} ice \citep{2012ApJ...758...38K}. 
While \citet{2023NatAs...7..431M} decomposed the \micron{4.27} absorption features into a combination of experimental profiles for \ch{H2O}-rich and \ch{CO}-mixed \ch{CO2} ices, we measured the composition of pure \ch{CO2} ice in addition to those ice mixtures based on the \micron{15} absorption feature for Ced 110 IRS4.

The presence of \ch{CO} gaseous absorption showing a broad wing structure around the \ch{CO} ice absorption suggests that the fitted gaseous profiles of warm \ch{CO}, higher than 70 K \citep{1990ApJ...363..554M}, indicate high thermal conditions in the protostellar envelope for most of our targets.
\citet{2022ApJ...935..137K} noted that the timescale since the last accretion burst for Per-emb 25 is estimated to be less than 1000 yr, based on the ALMA survey of \citet{2019ApJ...884..149H}.
This survey detected the peak radii of \ch{N2H+} and \ch{HCO+} emissions toward the protostars in the Perseus region and derived the timescale after the last accretion burst by considering the chemical interaction between the evaporated \ch{CO} and \ch{H2O} ice components and the subsequent destruction of \ch{N2H+} and \ch{HCO+} gases.
For Ced 110 IRS4, previous studies have noted its outflow activities and detected surrounding warm gaseous components of \ch{HCO+} \citep{2009A&A...508..259V,2020A&A...633A.126B}.
Specifically, RNO 91 has been reported as a protostar that recently underwent an accretion burst \citep{2021ApJ...919..116Y}.
Furthermore, the higher abundance of \ch{NH4+} in these targets, compared to its low abundance in B1-a, suggests high thermal conditions that can trigger UV photolysis of the \ch{NH3}-containing ice mixture, leading to the formation of \ch{NH4+} in the UV-irradiated protostellar envelope \citep{2003A&A...398.1049S}.

\section{Summary}
We comprehensively analyzed ice using combined AKARI IRC and Spitzer IRS spectra from four protostars at different evolutionary stages. 
To minimize silicate absorption, we employed a silicate-subtracted spectrum during the continuum determination process.

Our analysis involved fitting experimental ice absorbance data to identify intrinsic absorption profiles for various ice components. 
We used relevant ice profiles to fit the major absorption features of \ch{H2O}, \ch{CO2}, \ch{CO}, \ch{NH3}, and \ch{CH3OH}, including a combination of amorphous and crystalline \ch{H2O} ice profiles and \ch{H2O}-rich ice mixtures. 
We also considered additional ice components such as \ch{CH3CH2OH} and employed \ch{NH4+}, \ch{HCOOH}, and \ch{CH3CHO} ice profiles for specific absorption features at the mid-infrared range.

By deriving column densities, we found that the abundance of \ch{CO} ice decreases as protostellar targets evolve from Class 0 to Class \RomanNumeralCaps{1} due to the CO ice sublimation by the heating during protostellar mass accretion. 
To check the ice evolution in a given molecular cloud before and after star formation, we compared ice abundances between the protostar, Ced 110 IRS4 from this study and two dense cloud regions from the JWST observations in the Chameleon \RomanNumeralCaps{1} cloud.

The broad wing structure of \ch{CO} gaseous absorption around \ch{CO} ice indicates high thermal conditions in the protostellar envelope for Per-emb 25, Ced 110 IRS4, and RNO 91, with warm \ch{CO} temperatures higher than 70 K. 
Additionally, the higher abundance of \ch{NH4+} in these targets, compared to its low abundance in B1-a, suggests UV photolysis of \ch{NH3}-containing ice mixtures under high thermal conditions, forming \ch{NH4+} in the protostellar envelope.

\begin{table*}[t]
\centering
\caption{Source Sample of Protostellar Targets\label{tab:1}}
\begin{tabular}{lrrrrrrrrrr}
\toprule
Source & \multicolumn{2}{c}{Position}& Class & $L_{bol}$ & $T_{bol}$ & Distance & \multicolumn{2}{c}{AKARI} & \multicolumn{2}{c}{Spitzer} \\
\cline{2-3} \cline{8-9} \cline{10-11}  
       & $\alpha_{J2000.0}$ & $\delta_{J2000.0}$ & & ($L_{\odot}$) & (K) & (pc) & ObsID & Mode & ObsID & Module \\
\midrule
Per-emb 25 & 3 26 37.51 & 30 15 27.79 & 0$^{\rm a}$ & 1.1$^{\rm d}$ & 64$^{\rm d}$ & 293$^{\rm e}$ & 3470016 & Np & 9835520 & SL, LL \\
Ced 110 IRS4 & 11 06 46.44 & -77 22 32.20 & 0/I$^{\rm b}$ & 0.8$^{\rm f}$ & 56$^{\rm f}$ & 125$^{\rm f}$ & 1640190 & Np & 12692224 & SL, LL\\
B1 a & 3 33 16.67 & 31 07 54.87 & I$^{\rm a}$ & 2.2$^{\rm e}$ & 100$^{\rm e}$ & 293$^{\rm e}$ & 1640045 & Nc & 15918080 & SL, LL \\
RNO 91 & 16 34 29.32 & -15 47 01.4 & I$^{\rm c}$/II$^{\rm b}$ & 2.6$^{\rm f}$ & 340$^{\rm f}$ & 125$^{\rm f}$ & 3470013 & Np & 5650432 & SL, LH \\
\bottomrule

\end{tabular}
\tabnote{For the case of RNO 91, short-exposure data of the AKARI IRC were used to avoid the flux saturation around the \micron{3} absorption feature. The bolometric luminosity and temperature of the targets and their distances are taken from $^{\rm d}$ \citet{2008ApJS..179..249D}, $^{\rm e}$ \citet{2019ApJ...884..149H}, and $^{\rm f}$ \citet{2015ApJS..217...30L}. The slit sizes of the IRS SL, LL, and LH modules are ${3.}^{\prime\prime}6\times{57}^{\prime\prime}$, ${10.}^{\prime\prime}5\times{168}^{\prime\prime}$, and ${11.}^{\prime\prime}1\times{22.}^{\prime\prime}3$ with a pixel scale of ${1.}^{\prime\prime}8$, ${5.}^{\prime\prime}1$, and ${4.}^{\prime\prime}5$, respectively \citep{2004ApJS..154...18H}.\\
$^{\rm a}$ \citet{2009ApJ...692..973E} \\
$^{\rm b}$ \citet{2013A&A...554A..34I} \\
$^{\rm c}$ \citet{2009ApJ...705.1160C} \\
}
\end{table*}

\begin{table*}[t]
\centering
\caption{Best fitting laboratory data\label{tab:2}}
\setlength{\tabcolsep}{0.05in}
\begin{tabular}{ccccccccc}
\toprule
Mixture & & T   & & Peak Position & & Vibration Mode & & $A$$^{\rm a}$ \\
        & & (K) & & ($\mu$m)      & &                & & (10$^{-17}$ cm molecule$^{-1}$) \\

\midrule
\ch{H2O}:\ch{CO}:\ch{NH3}:\ch{CO2}:\ch{CH3OH} && 10 && 3.05 && \ch{H2O} stretch && 20 \\
$=$ 100:20:20:14:10                           &&    && 6.05 && \ch{H2O} bend && 1.0 \\
                                              &&    && 13.5 && \ch{H2O} libration && 3.0 \\
                             &&    && 4.67 && \ch{CO} stretch && 1.1 \\
                             &&    && 2.96 && \ch{NH3} stretch && 2.3 \\
                             &&    && 9.0 && \ch{NH3} umbrella && 1.7 \\
                             &&    && 4.27 && \ch{CO2} stretch && 7.6 \\
                             &&    && 15.2 && \ch{CO2} bend && 1.2 \\
                             &&    && 9.75 && \ch{CH3OH} C$-$O stretch && 1.4 \\
                             
\ch{CO}:\ch{CO2} $=$ 100:70 && 10 && 4.67 && \ch{CO} stretch && 1.1 \\
                             &&    && 4.27 && \ch{CO2} stretch && 7.6 \\
                             &&    && 15.2 && \ch{CO2} bend && 1.2 \\

\ch{CH3OH}:\ch{CO2} $=$ 1:2 && 10 && 3.53 && \ch{CH3OH} C$-$H stretch && 0.8 \\
                             &&    && 6.8 && \ch{CH3OH} C$-$H deformation && 1.0 \\
                             &&    && 9.75 && \ch{CH3OH} C$-$O stretch && 1.4 \\    
                             &&    && 4.27 && \ch{CO2} stretch && 7.6 \\
                             &&    && 15.2 && \ch{CO2} bend && 1.2 \\
                         
\ch{CO}:\ch{CH3OH} $=$ 1:1 && 15 && 4.67 && \ch{CO} stretch && 1.1 \\
                               &&    && 3.53 && \ch{CH3OH} C$-$H stretch && 0.8 \\
                               &&    && 6.8 && \ch{CH3OH} C$-$H deformation && 1.0\\
                               &&    && 9.75 && \ch{CH3OH} C$-$O stretch && 1.4 \\ 

Pure \ch{H2O} && 160 && 3.10 && stretch && 20 \\
              &&     && 6.10 && bend    && 1.0 \\
              &&     && 12.2 && libration && 3.0 \\
Pure \ch{CO2} && 20 && 4.27 && stretch && 7.6 \\
              &&    && 15.2 && bend && 1.2 \\
Pure \ch{NH3} && 10 && 2.96 && stretch && 2.3 \\
              &&    && 9.3 && umbrella && 1.7 \\
Pure \ch{CH3OH} && 15 && 3.53 && \ch{CH3OH} C$-$H stretch && 0.8 \\
                &&    && 6.8 && \ch{CH3OH} C$-$H deformation && 1.0\\
                &&    && 9.7 && \ch{CH3OH} C$-$O stretch && 1.4 \\               
Pure \ch{HCOOH} && 10 && 5.85 && stretch && 6.7$^{\rm b}$ \\
Pure \ch{CH3CHO} && 15 && 5.81 && stretch && 3.0 \\
                 &&    && 8.91 && combination && 0.53 \\
Pure \ch{CH4} && 10 && 7.7 && deformation && 0.7 \\
Pure \ch{NH4+} && 12 && 6.79 && bend$^{\rm c}$ && 4.4$^{\rm d}$ \\
               && 80 && 6.85 && bend$^{\rm c}$ && 4.4$^{\rm d}$ \\
\bottomrule
\end{tabular}
\tabnote{Experimental profiles for ice absorbance of the mixture and pure components are referred from \citet{1996A&A...315L.341E} and \citet{2004AdSpR..33...14F}, respectively.\\
$^{\rm a}$ \citet{2022ApJ...935..133J} \\
$^{\rm b}$ \citet{1999A&A...343..966S} \\
$^{\rm c}$ \citet{2015ARA&A..53..541B} \\
$^{\rm d}$ \citet{2003A&A...398.1049S}}
\end{table*}

\begin{landscape}
\begin{table}[t]

\centering
\setlength{\tabcolsep}{0.02in}
\caption{Ice Column Densities\label{tab:3}}
\begin{tabular}{lccccccccccccccccccccccc}
\toprule
Source & N(\ch{H2O}) && N(\ch{H2O}) && \multicolumn{4}{c}{N(\ch{CO2})} && \multicolumn{3}{c}{N(\ch{CO})} & N(\ch{HCOOH}) & N(\ch{CH3CHO}) & N(\ch{CH4}) & N(\ch{NH4+}) & \multicolumn{2}{c}{N(\ch{NH3})} && \multicolumn{4}{c}{N(\ch{CH3OH})} \\
       & Stretch && Libration && \multicolumn{4}{c}{Bend} && \multicolumn{3}{c}{Stretch} & stretch & stretch & deformtion & bend & \multicolumn{2}{c}{umbrella} && \multicolumn{4}{c}{C$-$O stretch} \\
        & Pure && Pure && Pure & \ch{H2O} & \ch{CO} & \ch{CH3OH} && \ch{H2O} & \ch{CO2} & \ch{CH3OH} & Pure & Pure & Pure & Pure & Pure & \ch{H2O} && Pure & \ch{H2O} & \ch{CO} & \ch{CO2} \\

\midrule
Per-emb 25 & 90.09 && 82.88 && 2.91 & 13.85 & 3.14 & 10.19 && 14.35 & 4.40 & 0.20 & 3.89 & 2.99 & 2.73 & 10.44 & 7.60 & 14.93 && 2.61 & 1.20 & 0.18 & 7.88 \\
            & (11.17) && (10.16) && (0.67) & (0.81) & (1.16) & (0.72) && (1.19) & (0.36) & (0.17) & (1.49) & (1.12) & (2.26) & (1.65) & (9.47) & (9.53) && (1.21) & (1.68) & (1.83) & (1.68) \\
Ced 110 IRS4 & 43.14 && 40.85 && 1.94 & 5.43 & 0.60 & 6.11 && 5.34 & 0.85 & 1.20 & 2.07 & 0.60 & 5.14 & 4.15 & 6.29 & 4.95 && 0.98 & 0.19 & 1.09 & 4.73 \\
            & (2.17) && (7.22) && (0.41) & (0.99) & (0.61) & (0.87) && (0.76) & (0.12) & (0.61) & (1.40) & (1.04) & (2.25) & (1.62) & (4.17) & (4.29) && (0.67) & (0.57) & (0.85) & (0.75) \\ 
B1 a & 113.53 && 102.14 && 8.06 & 12.16 & 0.90 & 10.17 && 8.59 & 1.27 & 0.20 & 6.69 & 2.69 & 6.08 & 7.31 & 17.13 & 4.95 && 19.56 & 4.83 & 0.20 & 7.85 \\
            & (12.29) && (6.16) && (0.86) & (1.20) & (0.98) & (1.75) && (1.00) & (0.72) & (0.33) & (1.38) & (1.09) & (2.20) & (1.72) & (2.74) & (1.48) && (0.55) & (1.19) & (1.86) & (2.14) \\
RNO 91 & 35.56 && 33.22 && 3.20 & 3.97 & 0.42 & 2.77 && 3.80 & 0.59 & 0.20 & 2.68 & 0.30 & 2.11 & 5.34 & 3.29 & 4.08 && $\cdot$ & 0.76 & 0.18 & 2.19 \\
            & (18.89) && (7.92) && (0.81) & (2.54) & (1.27) & (2.22) && (1.07) & (0.37) & (0.22) & (1.76) & (1.31) & (2.79) & (2.17) & (6.76) & (4.25) && $\cdot$ & (2.55) & (0.42) & (3.65) \\

\bottomrule
\end{tabular}
\tabnote{All column densities are in 10$^{17}$ molecules cm$^{-2}$. Uncertainties in parentheses based on statistical errors in the spectra. The column density estimation of each ice component is performed for the absorption feature corresponding to the vibration mode listed in the second row. The third row describes each ice mixture that composes the relevant ice absorption features.}
\end{table}
\end{landscape}

\begin{table*}[t]
\centering
\setlength{\tabcolsep}{0.03in}
\caption{Ice Abundances$^{\rm a}$\label{tab:4}}
\begin{tabular}{lcccccccc}
\toprule
Source & X(\ch{CO2}) & X(\ch{CO}) & X(\ch{HCOOH}) & X(\ch{CH3CHO}) & X(\ch{CH4}) & X(\ch{NH4+}) & X(\ch{NH3}) & X(\ch{CH3OH}) \\

\midrule
Per-emb 25   & 0.36$\pm$0.05 & 0.23$\pm$0.03 & 0.05$\pm$0.02 & 0.04$\pm$0.01 & 0.03$\pm$0.03 & 0.11$\pm$0.02 & 0.27$\pm$0.12 & 0.14$\pm$0.03 \\
             & 0.19$^{\rm b}$$-$0.42$^{\rm c}$ & 0.30$^{\rm a}$ & & & & & & \\ 
Ced 110 IRS4 & 0.34$\pm$0.06 & 0.18$\pm$0.04 & 0.05$\pm$0.04 & 0.01$\pm$0.02 & 0.12$\pm$0.06 & 0.10$\pm$0.04 & 0.28$\pm$0.12 & 0.15$\pm$0.03 \\
B1 a         & 0.31$\pm$0.03 & 0.10$\pm$0.01 & 0.07$\pm$0.01 & 0.03$\pm$0.01 & 0.06$\pm$0.02 & 0.07$\pm$0.02 & 0.22$\pm$0.02 & 0.32$\pm$0.03 \\
             & 0.20$^{\rm d}$ & & 0.03$^{\rm e}$ & & & 0.08$^{\rm e}$ & 0.03$^{\rm f}$ & 0.02$^{\rm e}$ \\ 
RNO 91       & 0.31$\pm$0.13 & 0.14$\pm$0.06 & 0.08$\pm$0.06 & 0.01$\pm$0.04 & 0.06$\pm$0.08 & 0.16$\pm$0.07 & 0.22$\pm$0.22 & 0.09$\pm$0.11 \\
             & 0.30$^{\rm d}$$-$0.32$^{\rm b}$ & 0.21$^{\rm g}$$-$0.30$^{\rm b}$ & 0.03$^{\rm e}$ & & & 0.05$^{\rm e}$ & 0.05$^{\rm f}$ & 0.06$^{\rm e}$ \\
\bottomrule
\end{tabular}
\tabnote{The \micron{13} \ch{H2O} libration mode used for X(ice) determination. \\
$^{\rm a}$ X(ice) $=$ N(ice)/N(\ch{H2O}).\\
$^{\rm b}$ \citet{2022ApJ...935..137K}, $^{\rm c}$ \citet{2012ApJ...758...38K}, $^{\rm d}$ \citet{2008ApJ...678.1005P}, $^{\rm e}$ \citet{2008ApJ...678..985B}, $^{\rm f}$ \citet{2010ApJ...718.1100B}, $^{\rm g}$ \citet{2003A&A...408..981P}}
\end{table*}



\acknowledgments
This work was supported by the National Research Foundation of Korea (NRF) grant funded by the Korea government (MSIT; grant No. 2021R1A2C1011718) and RS-2024-00416859.
This work is based on observations with AKARI, a JAXA project with the participation of ESA.
This work is based in part on observations made with Spitzer Space Telescope, which was operated by the Jet Propulsion Laboratory, California Institute of Technology, under contract with the National Aeronautics and Space Administration.
This research has made use of the NASA/IPAC Infrared Science Archive, which is operated by the Jet Propulsion Laboratory, California Institute of Technology, under contract with the National Aeronautics and Space Administration.
This research has made use of the VizieR catalogue access tool, CDS, Strasbourg, France (DOI : 10.26093/cds/vizier). The original description of the VizieR service was published in 2000, A\&AS 143, 23.
We thank an anonymous reviewer for constructive and insightful comments that we believe improved this paper.




\bibliography{IR_Spec_accept}{}

\end{document}